\documentclass[preprint,3p,times]{elsarticle}
\usepackage{graphicx}
\usepackage{dcolumn}
\usepackage{bm}
\usepackage{bm}
\usepackage{mciteplus}
\usepackage{amsmath}
\usepackage{amsfonts,amssymb}
\usepackage{color}
\usepackage{cancel} 
\usepackage{enumerate,soul}
\usepackage{amssymb}
\usepackage[normalem]{ulem}

\begin{document}
\begin{frontmatter}

\title{Cosmology and Non-additive Entropy.}

\author{A. Mart\'inez-Merino$^a$}
\ead{aldo.martinez@academicos.udg.mx}
 \author{M. Sabido$^b$}%
\ead{msabido@fisica.ugto.mx}
\address{$^a$ Departamento de Ciencias Naturales y Exactas, CU Valles, Universidad de Guadalajara.\\ Carretera Guadalajara - 
 Ameca Km. 45.5,  C.P. 46600, Ameca, Jalisco, M\'exico.\\
  $^b$ Departamento  de F\'{\i}sica de la Universidad de Guanajuato,
 A.P. E-143, C.P. 37150, Le\'on, Guanajuato, M\'exico.}

\begin{abstract}
{Non-additive entropies have been proposed as alternatives to understanding the thermodynamics of Black Holes. Moreover, it has been suggested that the difficulties of quantization of gravity are an indication that gravity is an emergent phenomenon, and therefore an entropic formulation is warranted. In this work, we  explore
cosmologies that come from non-additive entropies.} 
We focus our study on the cosmologies derived from  entropies in superstatistics.
\end{abstract}

\end{frontmatter}

\section{Introduction}
{The formulation of General Relativity (GR) is one of the most outstanding achievements in theoretical physics. The current results from gravitational wave astronomy cement GR as the appropriate theory to describe the gravitational interaction. Although the open problem of dark energy and dark matter is compatible with GR (if one proposes new exotic sources of matter and energy), current observations do not discard alternative theories of gravity. Furthermore, considering that a complete quantum theory of gravity is still missing after many decades of research, one can strongly look at alternatives to GR. We can follow the traditional line of reasoning by considering gravity as a fundamental interaction and from, some fundamental principle write down the corresponding theory (i.e., $f(R)$ \cite{Sotiriou:2008rp}, massive gravity \cite{deRham:2014zqa}, Horndeski, etc.).}

{A recent approach to understanding GR's  incompatibility  with quantum mechanics is considering the gravitational interaction  as an emergent phenomenon. Starting with an entropy proportional to the area, in \cite{Jac}, Einstein's equations are derived, verifying that one can consider GR as an entropic force. More recently, there was a resurgence with Verlinde's ideas are presented in \cite{Ver}, where he claims that Newtonian gravity is an entropic force in the sense of the emergent forces present in the study of polymers. This approach is motivated by the ideas in holography and the area relation for the entropy of black holes.} Other proposals make use of the holographic principle, for instance, in \cite{vR}, invoking the laws of entanglement to derive Einstein equations. The main ingredient in these formulations is the Bekenstein-Hawking entropy.  
As presented in \cite{O1}, we can 
obtain corrections to Newtonian gravity from a non-extensive entropy. Also in \cite{PS}, the consequences on planetary motion
were studied. {Since these formulations of gravity have an entropic origin, in principle, one can propose modifications to  gravity by analyzing changes to the entropy area law. These  corrections can come from considering non-extensive entropies.}

In this regard, in \cite{Shey1} and \cite{Shey3}, considering a
logarithm correction to Bekenstein-Hawking entropy,
the authors derive the Friedmann equations. In the present paper, 
following \cite{CCH}, we explore modification to the Friedmann-Robertson-Walker
cosmology, using the non-extensive entropies proposed in \cite{O1}. One
relevant feature of these entropies is that they do not depend on any
parameter but solely on the probability (in contrast to Tsallis entropy, that has a free parameter). 

The organization of this letter is as follows. In the next section, we review 
how the Friedmann equations are obtained from the approach in
\cite{CCH}. In Section 3, we derive the Friedmann equations using Obreg\'on's entropies 
and  study the limit of diluted matter density and the concentrated matter density limit. 
The last section is devoted to conclusions and final remarks.


\section{FRW cosmology from Clausius equation}

{To study the
dynamics of the universe, we need the Friedmann equations and the continuity equation. Although not a singular space, we can define an (apparent) horizon for 
the Friedmann-Robertson-Walker metric. Using the Clausius equation
and a linear relationship between the entropy and the area, one can derive the Friedmann
equations \cite{cai}}. 

Let $\tilde{r}$ be 
the radial coordinate of the FRW metric, the 
apparent horizon is at the radius
\begin{equation}
\tilde{r}_A = \frac{1}{\sqrt{H^2 + k/a^2}}, \label{apphor}
\end{equation}
where  $H=\dot{a} / a$ is the Hubble parameter, the dot represents
derivation with respect to cosmic time $t$ and $k$ is the
spatial curvature. As always,  the continuity equation is still satisfied with ordinary matter,
\begin{equation}\label{conteq}
\dot{\rho} + 3H (\rho + p) = 0.
\end{equation}

Now let $T$ and $S$ be the temperature and 
entropy, respectively, as measured by an observer at the apparent horizon. Assume that
a quantity of $\delta Q$ of energy pass through the area of this 
surface, which is equal to the product $T d S$, this gives
\begin{equation}
\delta Q = A (\rho + p) H \tilde{r}_A dt, \label{energy1}
\end{equation}
where the area of the apparent horizon is $A = 4\pi \tilde{r}_A^2$. Taking the 
differential of $S_{BH} = \frac{A}{4G}$ and assuming an implicitly dependence on $t$ from
$\tilde{r}_A$,  we find \footnote{We are using units where, $c = k_B = \hbar = 1$.}

\begin{equation}
T d S_{BH} = -\frac{H \tilde{r}^3_A}{G} \left( \dot{H} - \frac{k}{a^2} \right) d t.
\end{equation}
Comparing this relation with Eq. (\ref{energy1}), we arrive at the second Friedmann equation
\begin{equation}
\dot{H} - \frac{k}{a^2} = -4\pi G (\rho + p).
\end{equation}
Moreover, using the continuity equation, we arrive at
\begin{equation}
\frac{4\pi G}{3} \dot{\rho} = \frac{1}{2} \frac{d}{d t} \left( H^2 + 
\frac{k}{a^2} \right).
\end{equation}
Finally, integrating, we obtain the  Friedmann equation
\begin{equation}
H^2 + \frac{k}{a^2} = \frac{8\pi G}{3} \rho,
\end{equation}
where the integration constant has been absorbed in the energy density term.

{As suggested in \cite{CCH}, this approach has the advantage that it can be 
generalized to other entropies. For example, including an appropriate term on the entropy area relationship,
an effective cosmological constant is obtained \cite{isaac, luis}.
We can also use non-extensive entropies\footnote{These entropies have been used in connection with black holes quasinormal modes \cite{MMnS}.}, in particular, we will consider Obreg\'on's
entropies.}

\section{FRW cosmology from non-additive entropies}

Non-additive entropies can be obtained from a  general framework termed
\textit{superstatistics}. These entropies are the result of 
considering alternative probability distributions. In \cite{O1}, the author finds  entropies that depend 
solely on the probability -- unlike Tsallis entropy, which has an independent parameter $q$. The $q$-parameter has
no definite value, as it can reflect diverse degrees of complexity of a given 
system. 

We begin with the functional form of Obregon's entropy in terms of the black hole entropy \cite{MOR},
\begin{equation}
S_+ = e^{S_{BH}} \left( 1 - e^{-S_{BH}e^{-S_{BH}}}\right) = S_{BH} + 
\sum_{n=2}^\infty \frac{(-1)^{n+1}}{n!} S_{BH}^n
e^{-(n-1)S_{BH}}.
\end{equation}
Following the methodology of the
previous section, next we take the differential of this entropy as a function 
of the apparent horizon area
\begin{equation}
d S_+ = \left( 1 + \sum_{n=2}^\infty \frac{(-1)^{n+1}}{(n-1)!} \left[ 1 - \left( 1 - \frac{1}{n} \right) \frac{A}{4 G} \right] 
\left( \frac{A}{4 G} \right)^{n-1} e^{-(n-1)\frac{A}{4 G}} \right) \frac{d A}{4 G}. \label{diffplus}
\end{equation}
From the differential of the area in terms of the radius of  the
apparent horizon $Td A = 4\dot{\tilde{r}}_A d t$,
we get  the derivative of the radius with respect to cosmic time 
\begin{equation}
\dot{\tilde{r}}_A = - \tilde{r}_A^3 H \left( \dot{H} - \frac{k}{a^2} \right).
\end{equation}
Considering that the energy that traverse the surface of the space region is the same as in Eq.(\ref{energy1}) and equating to $T d S_+$, we get 
\begin{equation}\label{Corr2nd}
4\pi G (\rho + p) = -F(A) \left( \dot{H} - \frac{k}{a^2} \right),
\end{equation}
where we defined $F(A)=  1 + \sum_{n=2}^\infty \frac{(-1)^{n+1}}{(n-1)!} \left[ 1 - \left( 1 - \frac{1}{n} \right) \frac{A}{4 G} \right] 
\left( \frac{A}{4 G} \right)^{n-1} e^{-(n-1)\frac{A}{4 G}}$. 
This function encodes the corrections  due to the non-extensive entropy.
We can rewrite this expression
terms of the Hubble parameter using $A = 4\pi \tilde{r}_A^2$ 
and Eq.(\ref{apphor}). From the continuity equation, 
Eq.(\ref{Corr2nd}) is rewritten as
\begin{equation}
\frac{4\pi G}{3} \dot{\rho} =  H \left( \dot{H} - \frac{k}{a^2} \right) F\left(\frac{4\pi}{x}\right)= F\left(\frac{4\pi}{x}\right) \frac{1}{2} \frac{d}{d t} \left( H^2 + 
\frac{k}{a^2} \right).
\end{equation}
where we have defined the variable
$x \equiv H^2 + \frac{k}{a^2}$. After integrating ($x$ still
depends on a derivative respect to $t$), we have the equation
\begin{eqnarray}
\frac{8\pi G}{3} \rho &=& \int \left\{ 1 + \sum_{n=2}^\infty
\frac{(-1)^{n+1}}{(n-1)!} \left[ 1 - \left( 1 - \frac{1}{n}\right)
\frac{\pi}{Gx}
\right] \left( \frac{\pi}{G x} \right)^{n-1} e^{-(n-1)\frac{\pi}{G x}} \right\}
d x \nonumber \\
&=& x + \sum_{n=0}^\infty \frac{(-1)^{n+1}}{(n+1)!} \int e^{-(n+1)\frac{\pi}{G x}}\left[ 1 - \left( 1 -
\frac{1}{n+2}\right) \frac{\pi}{Gx}
\right] \left( \frac{\pi}{G x} \right)^{n+1}  d x.
\end{eqnarray}
{We single out the term $n=0$, and after integration we find,}
\begin{equation}
\frac{8\pi G}{3} \rho = x + \frac{\pi}{G} \sum_{n=0}^\infty \frac{(-1)^n}{(n+2)!} \left\{ \left( \frac{\pi}{G x} \right)^{n} e^{-(n+1)\frac{\pi}{G x}} 
- \frac{2}{(n+1)^{n}} \Gamma \left(n, \frac{(n+1)\pi}{Gx} \right) \right\},
\end{equation}
where $\Gamma(s,x)$ is the incomplete gamma function.

To write the corrected  Friedmann equation, we write the scale factor in terms of the energy density. This is
accomplished using the Lagrange inversion
theorem \cite{WW}. This theorem states that we can invert the equation $y = x + f(y)$ in terms of 
the variable $x$, by the relation
\begin{equation}
y = x + \sum_{n=1}^\infty \frac{1}{n!} \frac{d^{n-1}}{d x^{n-1}} \{ f(x) \}^n.
\end{equation}
With the suitable substitutions and Lagrange inversion theorem, we find 
that (up to third order in the exponential function) the modified Friedmann equation is
\begin{eqnarray}\label{FriedSplus}
H^2 + \frac{k}{a^2} &=& \frac{8\pi G}{3} \rho \left\{ 1 - \sum_{n=0}^\infty \frac{(-1)^n}{(n+2)!} \left( \frac{\pi}{G} \frac{3}{8\pi G \rho} 
\right)^{n+1} 
e^{-(n+1)\frac{\pi}{G} \frac{3}{8\pi G \rho}} \right. \nonumber \\
 &+& \sum_{n=0}^\infty \frac{2 (-1)^n}{(n+2)! (n+1)^{n}} \left( \frac{\pi}{G} \frac{3}{8\pi G \rho} \right)
\Gamma \left(n, \frac{(n+1)\pi}{G} \frac{3}{8\pi G \rho} \right)  + \frac{\pi^3}{4 G^3} \left( \frac{3}{8\pi G \rho} \right)^3 \left[ 1 - \frac{2G}{\pi} \frac{8\pi G \rho}{3} \right] 
e^{-\frac{2\pi}{G} \frac{3}{8\pi G \rho}} \nonumber \\
&-& \left.  \left( \sum_{n=0}^\infty \frac{2 (-1)^n}{(n+2)! (n+1)^{n}} \Gamma \left(n, \frac{(n+1)\pi}{G} \frac{3}{8\pi G \rho} \right) \right) 
\times f_3(\rho) + \dots \right\},
\end{eqnarray}
where $f_3$ is a function of powers of the the exponential $e^{-\frac{\pi}{G} \frac{3}{8\pi G \rho}}$,
\begin{eqnarray}
f_3(\rho) &=& \frac{\pi^3}{2 G^3} \left( \frac{3}{8\pi G \rho} \right)^3 \left[ 1 - \frac{2G}{\pi} \frac{8\pi G \rho}{3} \right] 
e^{-\frac{\pi}{G} \frac{3}{8\pi G \rho}} - \frac{\pi^4}{3 G^4} \left( \frac{3}{8\pi G \rho} \right)^4 \left[ 1 - \frac{3 G}{2\pi} 
\frac{8\pi G \rho}{3} \right] e^{-\frac{2\pi}{G} \frac{3}{8\pi G \rho}} \nonumber \\
& & + \frac{\pi^5}{8 G^5} \left( \frac{3}{8\pi G \rho} \right)^5 \left[ 1 - \frac{4 G}{3 \pi} \frac{8\pi G \rho}{3} \right] 
e^{-\frac{3\pi}{G} \frac{3}{8\pi G \rho}} + \dots ,
\end{eqnarray}
the next terms on the expansion are exponentially suppressed and therefore are neglected. 

{From the  modified Friedmann equation we focus our attention in two interesting limits. The first case of interest is for a highly dense Universe ($\rho \rightarrow \infty$), essentially the
conditions at the beginning of the Universe. 
The second case is for a diluted Universe ($\rho \rightarrow 0$), and corresponds to the late time evolution.}

For the first case, let us start by setting $\rho =1/\tilde{\rho}$ and consider the limit $\tilde{\rho} \rightarrow 0$. In this limit, the 
exponential functions go to unity, also the terms that are  powers of $\tilde{\rho}$ can be discarded. Moreover, in this limit
the incomplete gamma function goes to the ordinary gamma function  ($\Gamma(n, x) \rightarrow \Gamma(n)$). Thus, the 
relevant contributions come form the second and third terms in Eq.(\ref{FriedSplus}). Therefore, in this approximation we have
\begin{eqnarray}
H^2 + \frac{k}{a^2} &=& \frac{8\pi G}{3} \rho \left\{ 1 - \frac{1}{2} \frac{\pi}{G} \left( \frac{3}{8\pi G \rho} \right) 
+ \frac{\pi}{G} \left( \frac{3}{8\pi G \rho} \right) \sum_{n=1}^\infty \frac{(-1)^n}{(n+2)!} 
\frac{2}{(n+1)^{n}} \Gamma (n) \right\} \nonumber \\
&=& \frac{8\pi G}{3} \rho \left\{ 1 - \frac{\pi}{G} \left( \frac{3}{8\pi G \rho} \right) \left[ \frac{1}{2} -
\sum_{n=1}^\infty \frac{(-1)^n}{(n+2)!} \frac{2}{(n+1)^{n}} \Gamma (n) \right] \right\}.
\end{eqnarray}
We have to take caution with the term $n = 0$ in Eq.(\ref{FriedSplus}), since the function $\Gamma(0, x)$ diverges as $x$ approaches to zero. However, 
the divergence is logarithmic for $\rho\to\infty$, the quotient $\Gamma(0, \frac{1}{\rho})/\rho\to0$. 
 Finally we arrive to the Friedmann equation for this case, which is 
\begin{eqnarray}\label{early}
H^2 + \frac{k}{a^2} &=& \frac{8\pi G}{3} \rho \left\{ 1 - \frac{\pi}{G} \left( \frac{3}{8\pi G \rho} \right) \left[ \frac{1}{2}  
- \sum_{n=1}^\infty \frac{(-1)^n}{(n+2)!} \frac{2 (n-1)!}{(n+1)^{n}} \right] \right\} \nonumber \\
&=& \frac{8\pi G}{3} \rho \left\{ 1 - \frac{\pi}{G} \left( \frac{3}{8\pi G \rho} \right) \left[ \frac{1}{2} - 2 \sum_{n=1}^\infty \frac{(-1)^n}
{n(n+2)(n+1)^{n+1}} \right] \right\}.
\end{eqnarray}
The the sum on the l.h.s  converges\footnote{The value of the sum was calculated numerically.} $A = -0.157903$ and we rewrite Eq.(\ref{early}) as
\begin{equation}
H^2 + \frac{k}{a^2} = \frac{8\pi G}{3} \rho \left\{ 1 - \frac{3}{8 G^2 \rho} \left[ \frac{1}{2} + A \right] \right\}.
\end{equation}

For the second case ($\rho \rightarrow 0$), we see that the exponential function vanishes, and therefore all the terms of the series are suppressed  since all the extra terms are exponentially suppressed, and Eq.({\ref{FriedSplus}}) reduces to
the usual Friedmann equation. Therefore we conclude that the effects of the non-additive entropy are only relevant in the early Universe.


There exists two other entropies that only depend on the probability and do not have free parameters. This are known as $S_-$ and $S_\pm$.
As in the previous case, we can write their functional form in terms of the entropy $S_{BH}$, being 
\begin{eqnarray}
S_- &=& e^{S_{BH}} \left( e^{S_{BH} e^{-S_{BH}}} - 1 \right) = S_{BH} + \sum_{n=2}^\infty \frac{1}{n!} S_{BH}^n e^{-(n-1) S_{BH}}, 
\label{Obregontropy2}\\
S_\pm &=& S_+ + S_- = S_{BH} + \sum_{n=1}^\infty \frac{1}{(2n+1)!} S_{BH}^{2n+1} e^{-2n S_{BH}}. \label{Obregontropy3}
\end{eqnarray}
For these entropies we will follow the same procedure and definitions as before.

The differential for $S_-$ is
\begin{eqnarray}
d S_- &=& \left( 1 + \sum_{n=2}^\infty \frac{1}{(n-1)!} \left[ 1 - \left( 1 - \frac{1}{n} \right) \frac{A}{4 G} \right] 
\left( \frac{A}{4 G} \right)^{n-1} e^{-(n-1)\frac{A}{4 G}} \right) \frac{d A}{4 G} \nonumber \\
&=& F_-(A) \frac{4}{T} \dot{\tilde{r}}_A \frac{d t}{4 G},
\end{eqnarray}
where $F_-(A)$ is given by the terms in parenthesis on the r.h.s. of the previous expression.
From the definition of the apparent radius, Eq. (\ref{apphor}), after equating with Eq. (\ref{energy1}), we arrive at the corrected first 
Friedmann equation
\begin{equation}
4\pi G (\rho + p) = F_-(A) \left( \dot{H} - \frac{k}{a^2} \right),
\end{equation}
and with the aid of the continuity equation, after performing the integration we have 
\begin{equation}
\frac{8\pi G}{3} \rho = x + \frac{\pi}{G} \sum_{n=0}^\infty \frac{1}{(n+2)!} \left\{ \left( \frac{\pi}{G x} \right)^{n} e^{-(n+1)\frac{\pi}{G x}} 
- \frac{2}{(n+1)^{n}} \Gamma \left(n, \frac{(n+1)\pi}{Gx} \right) \right\}.
\end{equation}
Solving for $x$ in terms of the energy density $\rho$ and the inversion theorem, we find the corrected  Friedmann 
equation
\begin{eqnarray}
H^2 + \frac{k}{a^2} &=& \frac{8\pi G}{3} \rho \left\{ 1 - \sum_{n=0}^\infty \frac{1}{(n+2)!} \left( \frac{\pi}{G} \frac{3}{8\pi G \rho} \right)^{n+1} 
e^{-(n+1)\frac{\pi}{G} \frac{3}{8\pi G \rho}} \right. \nonumber \\
& & + \sum_{n=0}^\infty \frac{2}{(n+2)! (n+1)^{n}} \left( \frac{\pi}{G} \frac{3}{8\pi G \rho} \right)
\Gamma \left(n, \frac{(n+1)\pi}{G} \frac{3}{8\pi G \rho} \right) \nonumber \\
& & + \frac{\pi^3}{4 G^3} \left( \frac{3}{8\pi G \rho} \right)^3 \left[ 1 - \frac{2G}{\pi} \frac{8\pi G \rho}{3} \right] 
e^{-\frac{2\pi}{G} \frac{3}{8\pi G \rho}} \nonumber \\
& & \left. - \left( \sum_{n=0}^\infty \frac{2}{(n+2)! (n+1)^{n}} \Gamma \left(n, \frac{(n+1)\pi}{G} \frac{3}{8\pi G \rho} \right) \right) \times g_3(\rho) + \dots \right\},
\end{eqnarray}
where $g_3(\rho)$ is the function
\begin{eqnarray}
g_3(\rho) &=& \frac{\pi^3}{2 G^3} \left( \frac{3}{8\pi G \rho} \right)^3 \left[ 1 - \frac{2G}{\pi} \frac{8\pi G \rho}{3} \right] 
e^{-\frac{\pi}{G} \frac{3}{8\pi G \rho}} + \frac{\pi^4}{3 G^4} \left( \frac{3}{8\pi G \rho} \right)^4 \left[ 1 - \frac{3 G}{2\pi} 
\frac{8\pi G \rho}{3} \right] e^{-\frac{2\pi}{G} \frac{3}{8\pi G \rho}} \nonumber \\
& & + \frac{\pi^5}{8 G^5} \left( \frac{3}{8\pi G \rho} \right)^5 \left[ 1 - \frac{4 G}{3 \pi} \frac{8\pi G \rho}{3} \right] 
e^{-\frac{3\pi}{G} \frac{3}{8\pi G \rho}} + \dots.
\end{eqnarray}
Analyzing the limit $\rho \rightarrow \infty$, we look for a critical value in
the energy density. Considering the same arguments as before, the
equation we arrived after taking such limit is
\begin{equation}
H^2 + \frac{k}{a^2} = \frac{8\pi G}{3} \rho \left\{ 1 - \frac{3}{8 G^2 \rho} \left[ \frac{1}{2} - 2 \sum_{n=1}^\infty \frac{1}{n(n+2)(n+1)^{n+1}} 
\right] \right\},
\end{equation}
the the sum converges to the approximate value $B = 0.176475$. 

Finally, for the entropy $S_\pm$, the 
differential is
\begin{eqnarray}
d S_\pm &=& \left( 1 + \sum_{n=1}^\infty \frac{1}{(2n)!} \left[ 1 - \frac{2n}{2n+1} \frac{A}{4G} \right] \left( \frac{A}{4G} \right)^{2n} 
e^{-2n \frac{A}{4G}} \right) \frac{d A}{4 G} \\
&=& F_\pm(A) \frac{4}{T} \dot{\tilde{r}}_A \frac{d t}{4 G},
\end{eqnarray}
where $F_\pm(A)$ is given by the first term on the l.h.s. of the previous expression. After performing the correspondent steps, we arrive at the corrected 
Friedmann equation 
\begin{equation}
4\pi G (\rho + p) =F_\pm(A) \left( \dot{H} - \frac{k}{a^2} \right). 
\end{equation}
Performing the integration, we arrive to
\begin{equation}
\frac{8\pi G}{3} \rho = x - \frac{\pi}{G} \sum_{n=1}^\infty \frac{1}{(2n+1)!} \left\{ \left( \frac{\pi}{G x} \right)^{2n-1} e^{-\frac{2n \pi}{G x}}
- \frac{2}{(2n)^{2n-1}} \Gamma \left( 2n-1, 2n \frac{\pi}{G x} \right) \right\}.
\end{equation}
Solving for the scale factor  in terms of the energy density, we get the corrected Friedmann equation for this entropy,
\begin{eqnarray}
H^2 + \frac{k}{a^2} &=& \frac{8\pi G}{3} \rho \left\{ 1 + \sum_{n=1}^\infty \frac{1}{(2n+1)!} \frac{\pi^{2n}}{G^{2n}} \left( \frac{3}{8\pi G \rho} 
\right)^{2n} e^{-\frac{2n \pi}{G} \frac{3}{8\pi G \rho}} \right. \nonumber \\
& & - \frac{\pi}{G} \frac{3}{8\pi G \rho} \sum_{n=1}^\infty \frac{1}{(2n+1)!} \frac{2}{(2n)^{2n-1}} \Gamma \left( 2n-1, 2n \frac{\pi}{G}
 \frac{3}{8\pi G \rho} \right)
 \nonumber \\
& & \left. - \sum_{n=1}^\infty \frac{1}{(2n+1)!} \frac{2}{(2n)^{2n-1}} \Gamma \left( 2n-1, 2n \frac{\pi}{G} \frac{3}{8\pi G \rho} \right)
\times h_3(\rho) + \dots \right\},
\end{eqnarray}
the function $h_3(\rho)$ is given by
\begin{eqnarray}
h_3(\rho) &=& 2 \frac{\pi^5}{G^5} \left( \frac{3}{8\pi G \rho} \right)^5 \left[ 1 - \frac{G}{\pi} \frac{8\pi G \rho}{3} - \frac{G^2}{\pi^2}
\left( \frac{8\pi G \rho}{3} \right)^2 \right] e^{-\frac{2\pi}{G} \frac{3}{8\pi G \rho}} \nonumber \\
& & + \frac{1}{5} \frac{\pi^7}{G^7} \left( \frac{3}{8\pi G \rho} \right)^7 \left[ 1 - \frac{G}{\pi} \frac{8\pi G \rho}{3} - \frac{1}{2} 
\frac{G^2}{\pi^2} \left( \frac{8\pi G \rho}{3} \right)^2 \right] e^{-\frac{4\pi}{G} \frac{3}{8\pi G \rho}} + \dots .
\end{eqnarray}
Once again, taking the limit $\rho \rightarrow \infty$, we get
\begin{eqnarray}
H^2 + \frac{k}{a^2} &=& \frac{8\pi G}{3} \rho \left\{ 1 - \frac{\pi}{G} \frac{3}{8\pi G \rho} \sum_{n=1}^\infty \frac{1}{(2n+1)!} \frac{2}{(2n)^{2n-1}}
 \Gamma \left( 2n-1 \right) \right\} \nonumber \\
&=& \frac{8\pi G}{3} \rho \left\{ 1 - \frac{3}{4 G^2 \rho} \sum_{n=1}^\infty \frac{1}{4 n^2 - 1} \frac{1}{(2n)^{2n}} \right\}.
\end{eqnarray}
In this case we are not worried for any divergent terms. The sum has the approximated value of $C=0.0835944$. 


For these entropies in the case ($\rho \rightarrow 0$), we see that the exponential function vanishes, all the terms of the series are suppressed, and the Friedmann equation reduces to
the usual Friedmann equation. This behavior is the same for all of Obregon's entropies. Consequently, the effects of these entropies are only relevant in the early Universe and for the late time, evolution is indistinguishable from the usual cosmological description.

\section {Conclusions and suggestions for further research}

In this paper, we have explored the effects in the cosmology of non-additive entropies that only depend on probability.
These entropies, are known as Obregon's entropies, and have the peculiarity that there are no free parameters.
Using the Clausius equation to relate the energy contained in a region of
space with the entropy associated with the apparent horizon of that region,
we obtained corrected versions of the first and second Friedmann equations
using non-extensive entropies that do depend on the probability.
We have focused our attention  the early and late time Universe. These epochs correspond to the infinite density
and of diluted matter limits.
All the corrections are suppressed from the Friedmann equation in the diluted limit, and the Friedmann equation reduces to
the usual Friedmann equation (this behavior is the same for all of Obregon's entropies). Therefore, the effects are only relevant in the very early Universe. Consequently, one can conclude that non-extensive statistics modifies the dynamics of quantum cosmology. This subject is under research and will be reported elsewhere.

\section*{Acknowledgments}
The authors would like to thank O. Obreg\'on for comments for the present
manuscript. M. S. is supported by CIIC-071/2022.

\end{document}